\documentclass[sigconf,preprint]{acmart}

\AtBeginDocument{%
  }

\acmDOI{}
\acmISBN{}
\acmConference{}{}{}{}

\usepackage[most]{tcolorbox} 
\usepackage{CJKutf8}

\begin{document}

\title{Edit, But Verify: An Empirical Audit of Instructed Code-Editing Benchmarks}

\author{Amir M. Ebrahimi\textsuperscript{*}}
\affiliation{%
  \institution{Queen's University}
  \city{Kingston}
  \country{Canada}}
\email{amir.ebrahimi@cs.queensu.ca}

\author{Gopi Krishnan Rajbahadur\textsuperscript{*}}
\affiliation{%
  \institution{Queen's University}
  \city{Kingston}
  \country{Canada}}
\email{grajbahadur@acm.org}

\renewcommand{\shortauthors}{Ebrahimi and Rajbahadur}

\begin{abstract}
Instructed code editing, where an LLM modifies existing code based on a natural language instruction, accounts for roughly 19\% of real-world coding assistant interactions. Yet very few benchmarks directly evaluate this capability. From a survey of over 150 code-related benchmarks, we find that only two, CanItEdit and EDIT-Bench, target instructed code editing with human-authored instructions and test-based evaluation. We audit both by comparing their programming languages, edit intents, and application domains against distributions observed in the wild (Copilot Arena, AIDev, GitHub Octoverse), and by measuring test counts, statement coverage, and test scope across all 213 problems. Both benchmarks concentrate over 90\% of evaluation on Python while TypeScript, GitHub's most-used language, is absent. Backend and frontend development, which together constitute 46\% of real-world editing activity, are largely missing, and documentation, testing, and maintenance edits (31.4\% of human PRs) have zero representation. Both benchmarks have modest test counts (CanItEdit median 13, EDIT-Bench median 4), though CanItEdit compensates with near-complete whole-file coverage and fail-before/pass-after validation. 59\% of EDIT-Bench's low-coverage suites would not detect modifications outside the edit region. EDIT-Bench has 15 problems that are not solved by any of 40 LLMs and 11 of these problems trace failures to poor benchmark artifacts rather than model limitations. Further, 29\% of EDIT-Bench problems and 6\% of CanItEdit problems share a codebase with at least one other problem within the benchmark. In summary, these benchmarks measure a narrower construct than deployment decisions require. We therefore propose six empirically grounded desiderata and release all audit artifacts so the community can build instructed code-editing benchmarks whose scores reliably reflect real-world editing capability.
\end{abstract}



\maketitle

\let\thefootnote\relax\footnotetext{\textsuperscript{*}Both authors contributed equally to this work.}
\section{Introduction}
 \smallskip
\begin{flushright}
\begin{minipage}{0.78\linewidth}
\hrule
\vspace{0.6em}
\raggedleft
\itshape
``Add error handling for the API timeout case.''\\[0.35em]
{\small A developer at 3:15\,AM, highlighting code in their IDE and expecting an in-place edit.}
\vspace{0.6em}
\hrule
\end{minipage}
\end{flushright}

Instructed code editing~\cite{cassano2024canitedit} accounts for roughly 19\% of real-world coding assistant conversations~\cite{cassano2024canitedit} and sits behind much of the work developers hand to AI: feature implementation, bug fixing, refactoring, test authoring, and documentation updates~\cite{sergeyuk2025}. These benchmarks increasingly serve as real-world proxies for deciding if an LLM can be used as part of AIware like Cursor or GitHub Copilot. For these benchmarks to act as real-world proxies, two key things must be true: the benchmark must test the edits developers actually make in the real world, and the test oracles must be strong enough to back up what a passing score is taken to mean. When either condition fails, the scores do not just look incomplete. They point deployment decisions in the wrong direction.
 
To study this validity question, we surveyed over 150 code-related benchmarks provided in the recent survey by Yang~et~al.~\cite{yang2025codesurvey} and filtered for human-authored natural language instructions on existing code, single-file IDE-local scope, and test-based evaluation. Two benchmarks survived this filtering: CanItEdit~\cite{cassano2024canitedit} and EDIT-Bench~\cite{chi2025editbench}. We then empirically audit if these two benchmarks can act as real-world proxies for evaluating the instructed code-editing capability of an LLM through the following two research questions.
 
\textbf{RQ1: How representative are CanItEdit and EDIT-Bench of real-world AI-assisted code editing?}
A benchmark cannot serve as a real-world proxy if it tests the wrong capabilities. We compared both benchmarks against Copilot Arena~\cite{chi2025copilotarena}, the AIDev dataset~\cite{li2025aidev}, and GitHub Octoverse~\cite{github2025octoverse} across programming language, edit intent, and application domain. Over 90\% of problems across both benchmarks target Python alone, a language that accounts for 20--30\% of real-world activity, while TypeScript, now GitHub's most-used language, is entirely absent. Both benchmarks concentrate 79--86\% of problems on feature and fix tasks; four categories representing 31.4\% of human-authored PRs (documentation, testing, build/CI, maintenance) have zero representation. Domain coverage is similarly skewed: CanItEdit devotes 68.6\% of problems to algorithm design (18\% real-world share); EDIT-Bench allocates 36.1\% to AI/ML (7\% real-world share). Backend and frontend development, which together account for 46\% of real-world editing activity, are absent from CanItEdit and reach only 24.1\% combined in EDIT-Bench.
 
\textbf{RQ2: How adequate are the CanItEdit and EDIT-Bench tests as correctness oracles for code editing?}
Even a representative benchmark produces unreliable scores if its tests cannot distinguish correct edits from subtly broken ones. To see if the tests provided in these benchmarks are adequate, we executed every test suite in instrumented Docker environments and measured statement coverage at both whole-file and diff-region granularity. CanItEdit compensates for modest test counts (median 13) with near-complete whole-file coverage (median 100\%) and fail-before/pass-after validation. EDIT-Bench's suites are thinner: a median of 4 tests per problem, whole-file coverage of 40\%, and 14 problems relying on a single test as the entire correctness signal. More critically, 59\% of low-coverage suites would not detect extraneous modifications outside the requested change, an important concern regarding the validity of EDIT-Bench since LLM-powered editing tools routinely modify code beyond the requested scope in practice~\cite{mozannar2025transformcode}. Furthermore, Liu~et~al.~\cite{liu2023evalplus} show that code-related benchmarks with less than 10 tests per problem routinely overestimate the functional correctness of the generated code. In summary, both benchmarks answer \emph{was the edit made?} more reliably than \emph{was only the requested edit made?}.
 
Benchmark artifacts and code duplication further erode the evaluation surface that RQ1 and RQ2 already found wanting. Of the 15 EDIT-Bench problems no model solves, 11 (73\%) trace to benchmark artifacts (infrastructure bugs, impossible test assertions, and unstated environmental requirements) rather than model limitations, inflating the apparent difficulty ceiling. Code duplication further reduces evaluation diversity: 29\% of EDIT-Bench problems share a codebase with at least one other problem, leaving only 86 genuinely distinct code contexts out of 108 (CanItEdit exhibits 103 distinct contexts out of the 105). Where RQ1 shows that these benchmarks do not reflect real-world usage, RQ2 shows the test oracles have serious limitations. Together, these findings imply that current instructed code-editing benchmarks are not a reliable proxy for measuring the code-editing performance of LLMs.
 
We present our paper as a systematic empirical audit of benchmark validity rather than a new benchmark proposal. Based on our findings, we outline six empirically grounded desiderata for constructing instructed code-editing benchmarks whose scores can reliably serve as real-world proxies (Section~\ref{sec:implications}). Key contributions of our paper are:
\begin{itemize}
  \item An \textbf{empirical audit} of CanItEdit and EDIT-Bench across five dimensions (programming language, edit intent, application domain, test suite adequacy, and data completeness), grounded in external real world reference distributions~\cite{chi2025copilotarena,li2025aidev,github2025octoverse}.
  \item A \textbf{diagnosis of benchmark artifacts}: A deep dive into universally-unsolved problems in EDIT-Bench and code duplication patterns across both the benchmarks that affect difficulty profiles and evaluation diversity.
  \item \textbf{Six empirically grounded desiderata} for next-generation instructed code-editing benchmarks, each tied to a specific audit finding.
\end{itemize}
 
\smallskip\noindent\textbf{Data Availability.} All the prompts and the results of the manual analysis used in our paper are provided in the appendix. 

\smallskip\noindent\textbf{Paper Organization.} Section~\ref{sec:methodology} describes our benchmark selection criteria and the two studied benchmarks. Section~\ref{sec:rqs} presents the results of RQ1 (representativeness) and RQ2 (test adequacy). Section~\ref{sec:discussion} analyzes universally unsolved problems
and code duplication. Section~\ref{sec:implications} proposes six desiderata for next-generation instructed code-editing benchmarks. Sections~\ref{sec:threats} and~\ref{sec:related-work} discuss threats to validity and related work.

\section{Methodology}\label{sec:methodology}
We audit CanItEdit and EDIT-Bench along two axes: whether they test representative edits (RQ1) and whether their test oracles are strong enough to certify correctness (RQ2). This section describes how we selected these benchmarks and summarizes their design.

\subsection{Benchmark Selection}
To systematically identify the benchmarks studied in this work, we began with the 180 unique benchmarks cataloged by Yang~et~al.~\cite{yang2025codesurvey}, spanning code generation, editing, efficiency, reasoning, and agentic tasks, and supplemented this with a targeted search that surfaced EDIT-Bench~\cite{chi2025editbench} as an additional candidate not yet included in the survey. One author then applied progressive inclusion and exclusion criteria to this combined pool. The first filter retained only benchmarks related to code generation or code editing, reducing the set to 101 candidates. From these, we applied three further exclusion criteria: (1)~the benchmark must not be agentic or multi-turn, (2)~the benchmark must target code editing rather than code generation from scratch, and (3)~the benchmark must evaluate holistic editing tasks, not only bug fixing. Three benchmarks survived: CodeEditorBench~\cite{guo2025codeeditorbench}, CanItEdit~\cite{cassano2024canitedit}, and EDIT-Bench~\cite{chi2025editbench}. A fourth criterion, (4)~that editing instructions must be authored by humans rather than generated by an LLM, eliminated CodeEditorBench, whose instructions are LLM-generated, leaving CanItEdit and EDIT-Bench as the two benchmarks studied in this work.
 
\subsection{Studied Benchmarks}
 
\smallskip\noindent\textbf{CanItEdit.}
CanItEdit~\cite{cassano2024canitedit} comprises 105 hand-crafted Python code-editing problems created by eight experienced programmers with a designated lead reviewer. Each problem includes \textit{before} and \textit{after} code segments and two human-written instructions (a detailed \textit{descriptive} and a minimal \textit{lazy} variant), yielding 210 problem instances. Problems are evenly split across corrective, adaptive, and perfective edits (35 each), spanning data structures, algorithms, mathematics, and data science, with 22 requiring external libraries (NumPy, Pandas, PyTorch, Z3). Test suites employ unit tests, property-based testing, mocking, fuzzing, and integration tests, validated by a pipeline that enforces 100\% line coverage and fail-before/pass-after checks, ensuring each suite is sensitive to the intended edit. Evaluation uses pass@k in Docker containers alongside an ExcessCode metric for uncovered changed lines. The benchmark is exclusively Python and has been widely adopted to evaluate LLM code-editing capability~\cite{zhang2025generating,wei2024selfcodealign,lozhkov2024starcoder}.

\smallskip\noindent\textbf{EDIT-Bench.}
EDIT-Bench~\cite{chi2025editbench} sources its 108 core problems from Copilot Arena, a VS Code extension used by nearly 500 developers that captures authentic instructions and code contexts from daily coding tasks. EDIT-Bench was proposed as a corrective to benchmarks like CanItEdit, arguing that hand-crafted problems resemble LeetCode exercises and miss the diversity of real developer behavior. Each problem includes the user instruction, code context, a highlighted edit region, and cursor position, spanning Python and JavaScript. Original instructions are predominantly English (85) with Russian (15), Spanish (4), Chinese (3), and Polish (1); the authors expand to 540 problems via GPT-4o translation, but we use only the 108 core problems in the remainder of this study, since the rest are translation duplicates. Five experienced programmers manually created test harnesses, as raw extension data lacks test cases. Unlike CanItEdit, EDIT-Bench releases neither reference solutions nor fail-before/pass-after validation or coverage metadata. Problems span feature addition (43\%), modification (27\%), bug fixing (22\%), and optimization (8\%).

\section{Research Questions}~\label{sec:rqs}

\smallskip
\noindent\textbf{RQ1: How representative are CanItEdit and EDIT-Bench of real-world AI-assisted code editing?}

\smallskip
\noindent\textbf{Motivation.}
Instructed code-editing benchmarks can only produce meaningful evaluations if they reflect how developers actually use LLMs for code editing. Copilot Arena~\cite{chi2025copilotarena}, which collected over 4.5 million inline code completions from real IDE sessions, reports that developers employ LLMs for coding tasks across 103 programming languages, with no single language exceeding 30\% of activity. The three most common application domains are \textit{Backend and API Development}, \textit{Frontend Development and UI Design}, and \textit{Algorithm Design and Problem Solving}. Li~et~al.~\cite{li2025aidev} collected 456K+ agent-authored pull requests across 61K public repositories (AIDev  dataset) on GitHub and found that agents routinely contribute documentation (12.8\%), tests (4.5\%), refactoring (6.9\%), and maintenance tasks beyond new feature additions and bug fixes. Both Li~et~al.~\cite{li2025aidev} and the GitHub Octoverse 2025 report~\cite{github2025octoverse} note that TypeScript, JavaScript, Java, C\#, and Go are actively used alongside Python, with TypeScript surpassing Python as the most-used language by contributor count. The Octoverse report further documents that over 1.1 million public repositories now import an LLM SDK, representing 178\% year-over-year growth. These are the contexts in which LLMs are used for code editing. For benchmark scores to serve as a reliable proxy for real-world editing capability, benchmark tasks should encompass this diversity. We therefore audit the representativeness of CanItEdit and EDIT-Bench along three dimensions: \textit{programming language}, \textit{edit intent} (i.e., the type of edit e.g., test function writing, infrastructure code), and \textit{application domain}.

\smallskip\noindent\textbf{Approach.} For \textit{programming language}, we extract the language composition of CanItEdit~\cite{cassano2024canitedit} and EDIT-Bench~\cite{chi2025editbench} from their published repositories and compare against distributions from the three reference sources above. For \textit{edit intent}, we classify all benchmark tasks using the Conventional Commit taxonomy adopted by the AIDev dataset~\cite{li2025aidev} (\texttt{feat}, \texttt{fix}, \texttt{docs}, \texttt{test}, \texttt{refactor}, \texttt{perf}, \texttt{style}, \texttt{chore}, \texttt{build}, \texttt{ci}) and compare against AIDev task-type distributions for both human-authored and agent-authored PRs. For \textit{application domain}, we classify all benchmark tasks into the same seven domain categories used by Copilot Arena~\cite{chi2025copilotarena} and compare the resulting distribution against theirs. We further classify all AI/ML-tagged problems into three subcategories (Traditional ML Development, ML-Adjacent Tooling, and AI-Native Application Development) to assess whether the studied benchmarks represent emerging software development paradigms. All annotation tasks use Claude Sonnet 4.6 with the prompts provided in Appendix~\ref{fig:prompt-app-domain}, Appendix~\ref{fig:prompt-task-type}, and Appendix~\ref{fig:prompt-ml-subtype} as recent studies show that LLMs exhibit strong performance when annotating software engineering artifacts~\cite{ahmed2025can,li2025aidev}.

\smallskip\noindent\textbf{Results.}

\smallskip\noindent\textit{Programming Language.} \textbf{Both EDIT-Bench and CanItEdit are overwhelmingly Python-centric, with minimal to no coverage of other commonly used languages in AI-assisted development.} TypeScript, now the most-used language on GitHub and the dominant language across autonomous coding agents~\cite{github2025octoverse, li2025aidev}, is entirely absent from both benchmarks. CanItEdit is exclusively Python (105/105 problems)~\cite{cassano2024canitedit}. EDIT-Bench allocates 89.8\% of its instances to Python, with the remaining 10.2\% split between JavaScript and JavaScript/React, and zero coverage of Java, C\#, Go, or Rust~\cite{chi2025editbench}. Combined, the two benchmarks concentrate over 90\% of their tasks on a single language that accounts for roughly 20--30\% of real-world AI-assisted coding activity.

Such narrow language coverage undermines evaluation validity in two compounding ways. Benchmark results provide no signal for developers editing TypeScript, reviewing Java PRs from Codex, or debugging C\# with Cursor, since their languages are entirely absent from evaluation. Meanwhile, even the Python signal is likely inflated by training-data density: recent work shows LLMs disproportionately favour Python due to training-data saturation~\cite{twist2025llms}, and model performance on coding problems correlates with language popularity in training corpora rather than intrinsic task difficulty\footnote{\url{https://www.thetechpanda.com/benchmarking-llm-coding-proficiency-across-languages/43892/}}. High pass@1 on Python editing tasks may therefore reflect the ceiling of model comfort rather than a reliable indicator of real-world code editing capability.

\begin{figure}[t]
\centering
\includegraphics[width=1.05\columnwidth]{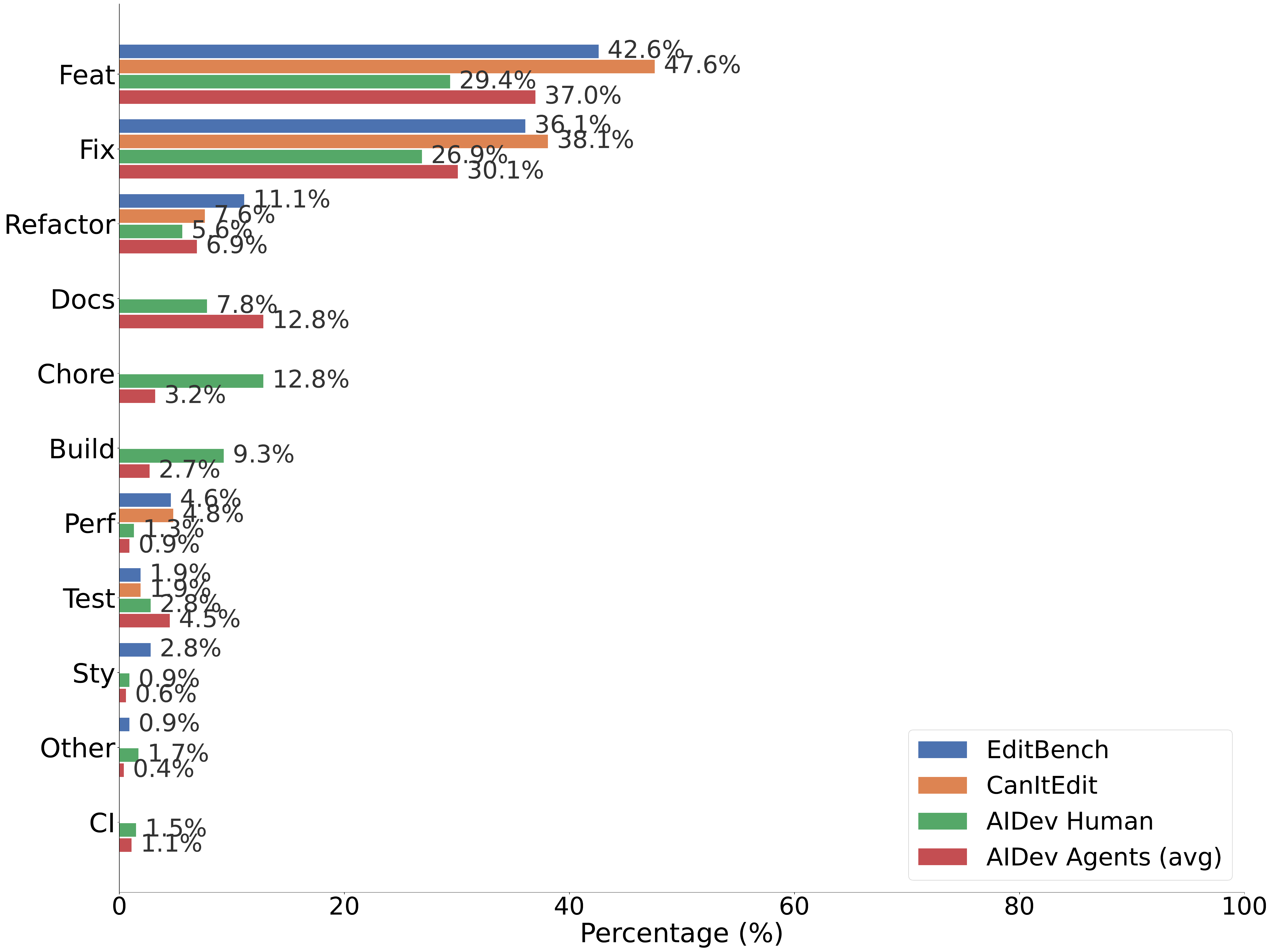}
\caption{Edit-intent distribution across benchmarks and real-world AI-assisted development (AIDev).}
\label{fig:edit-intent}
\end{figure}

\smallskip
\noindent\textit{Edit Intent.} \textbf{Both benchmarks concentrate overwhelmingly on new feature implementation and fix tasks, leaving entire categories of real-world editing activity unrepresented} (Figure~\ref{fig:edit-intent}). EDIT-Bench allocates 78.7\% of its problems to feat (42.6\%) and fix (36.1\%); CanItEdit allocates 85.7\% to the same two categories (47.6\% feat, 38.1\% fix). Four categories that collectively account for 31.4\% of human PRs and 19.8\% of agent PRs in AIDev have exactly zero representation in both benchmarks: docs, chore, build, and ci. Documentation alone represents 12.8\% of agent-authored PRs, making it the third-largest category in real-world agent activity, yet no benchmark problem exercises it. Test authoring is similarly underserved: agents devote 4.5\% of their PRs to test tasks, but each benchmark contains only two test-related problems (1.9\%). Conversely, both benchmarks overweight performance optimization (4.6--4.8\%) relative to real-world prevalence (0.9--1.3\%).

Benchmark scores therefore provide no signal on whether a model can update documentation in response to a code change, author a meaningful test case, refactor CI/build configurations, or perform routine project upkeep. These are not edge cases but account for nearly a third of real-world human editing activity and a fifth of agent editing activity~\cite{li2025aidev}. Their absence means current benchmarks evaluate code editing through a feature-and-fix lens that misses the breadth of software engineering work these models are actually used for.

\begin{figure}[t]
\centering
\includegraphics[width=\columnwidth]{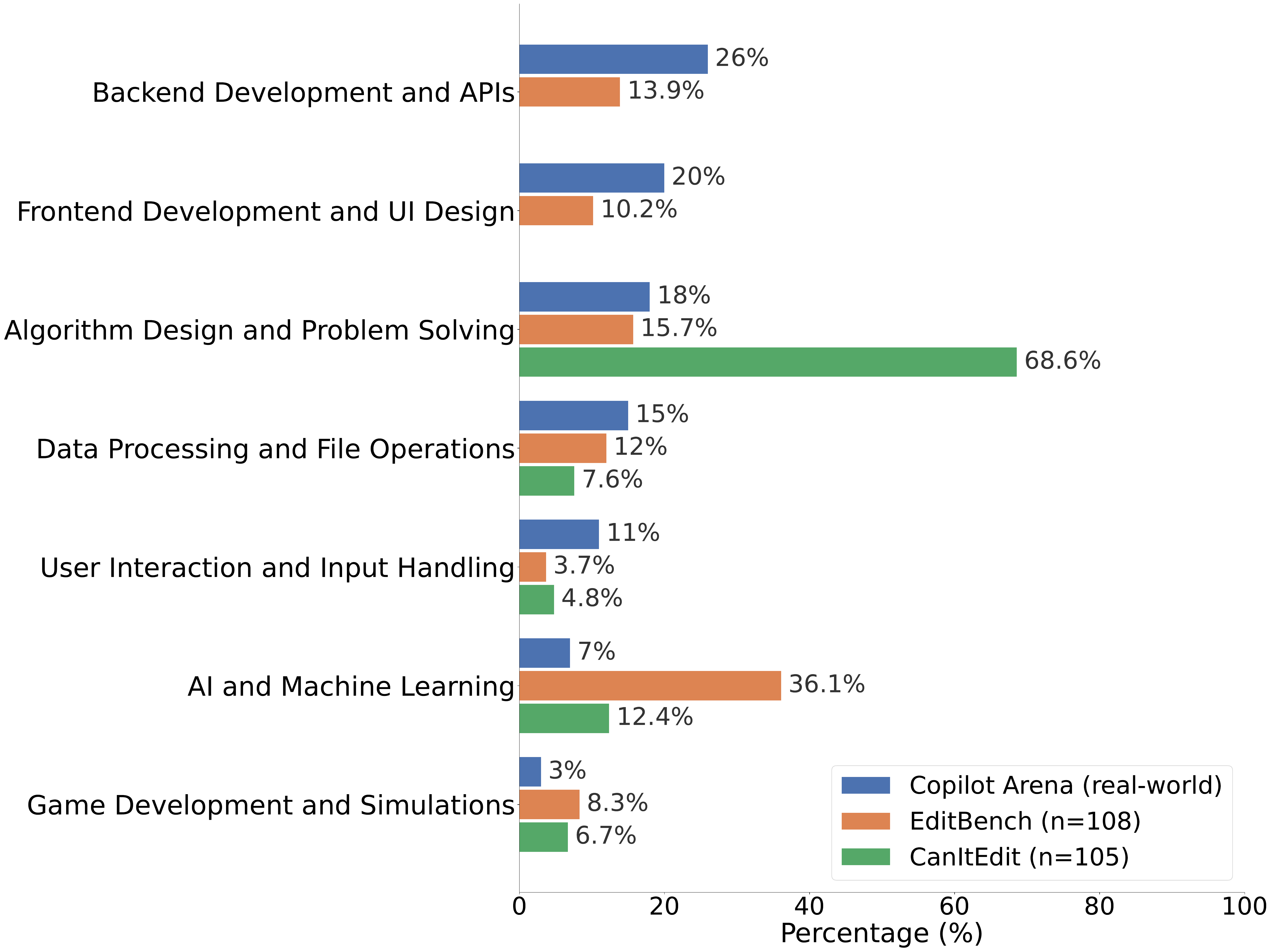}
\caption{Application-domain distribution across benchmarks and real-world AI-assisted editing (Copilot Arena).}
\label{fig:app-domain}
\end{figure}

\smallskip\noindent\textit{Application Domain.} \textbf{Both CanItEdit and EDIT-Bench substantially overrepresent AI/ML and algorithmic domains while underrepresenting the backend and frontend development work that dominates real-world AI-assisted editing} (Figure~\ref{fig:app-domain}). EDIT-Bench allocates 36.1\% of its problems to AI and Machine Learning, a 5-fold overrepresentation relative to its 7\% real-world prevalence in Copilot Arena. CanItEdit concentrates 68.6\% in Algorithm Design and Problem Solving, nearly four times its 18\% real-world share. Conversely, Backend Development (26\% in Copilot Arena) has zero coverage in CanItEdit and only 13.9\% in EDIT-Bench; Frontend Development (20\% in Copilot Arena) is likewise absent from CanItEdit and reaches only 10.2\% in EDIT-Bench. Both benchmarks achieve reasonable coverage for Data Processing (12.0\% and 7.6\% vs.\ 15\% in Copilot Arena) and Game Development (8.3\% and 6.7\% vs.\ 3\%), and EDIT-Bench's broader spread across domains reflects a more deliberate effort to capture domain diversity than CanItEdit's narrow algorithmic focus.

\begin{table}[t]
\centering
\caption{Distribution of AI/ML problems by subcategory.}
\label{tab:ml-subcategory}
\small
\begin{tabular}{@{}lrrrr@{}}
\toprule
\textbf{ML Subcategory} & 
\multicolumn{2}{c}{\textbf{EDIT-Bench (39)}} & 
\multicolumn{2}{c}{\textbf{CanItEdit (13)}} \\
\cmidrule(lr){2-3} \cmidrule(lr){4-5}
 & \textit{n} & \textit{\%} & 
   \textit{n} & \textit{\%} \\
\midrule
Traditional ML Dev        & 11 & 28.2 & 11 & 84.6 \\
ML-Adjacent Tooling       & 18 & 46.2 &  1 &  7.7 \\
AI-Native Application Dev & 10 & 25.6 &  1 &  7.7 \\
\bottomrule
\end{tabular}
\end{table}

Even within the overweighted AI/ML slice, the distribution of subcategories misses the most significant shift in modern software development 
(Table~\ref{tab:ml-subcategory}). CanItEdit concentrates 84.6\% of its AI/ML problems (11 of 13) in Traditional ML, with only a single problem targeting AI-native application development. EDIT-Bench is more balanced but still allocates only 25.6\% (10 of 39) to 
AI-native tasks, while placing nearly half (46.2\%) in ML-Adjacent Tooling. Despite LLM-integrated development being the fastest-growing category on GitHub, it is the least represented subcategory in both benchmarks.


\smallskip\noindent\textbf{RQ2: How adequate are the CanItEdit and EDIT-Bench tests as correctness oracles for code editing?}
 
\smallskip\noindent\textbf{Motivation.} Benchmark scores inherit the strength of the tests behind them. Liu~et~al.~\cite{liu2023evalplus} showed that commonly used code benchmarks often contain fewer than 10 tests per problem, that these tests are simplistic and fail to capture corner cases, and that strengthening HumanEval's tests lowered pass@k of several frontier models by 19.3--28.9\% and reversed model rankings. Subsequent work finds that code benchmarks often have poor overall coverage, with low branch coverage (74.4\%) and low mutation scores (87.7\%)~\cite{liu2025testadequacy}. Yet this issue is rarely surfaced: only 8.7\% of code benchmarks that use tests as the oracle explicitly report coverage metadata~\cite{cao2025how2bench}. Neither EDIT-Bench nor CanItEdit is among them. We therefore measure three observable properties of each benchmark's test suites: test count per problem, as a basic indicator of evaluation density; whole-file and diff-region statement coverage of the reference solutions, to assess how much of the code the tests actually exercise; and test scope, to determine whether suites verify only that the requested edit was applied or also check for unwanted modifications and regressions. If instructed code-editing benchmark scores are to serve as a proxy for real-world adoption, particularly that a model is suitable for AIware (for example as the default model in a code IDE), then a benchmark that checks whether an edit was made without checking for regressions provides a weaker signal than the score implies.
 
\smallskip\noindent\textbf{Approach.} We executed each problem's test suite in an isolated Docker environment instrumented with \texttt{coverage.py} for Python and Jest/Istanbul for JavaScript. For CanItEdit, reference solutions are available for all 105 problems, so we computed whole-file statement coverage directly. EDIT-Bench does not release reference solutions. However, the benchmark publishes per-problem pass rates for each of the 40 models evaluated in the original study~\cite{chi2025editbench}. For each problem, we identified models that achieved a perfect score across all tests and regenerated solutions from one such model until it produced a passing output. This yielded reference solutions for 93 of 108 problems; the remaining 15 had no model with a perfect pass rate and were excluded from coverage analysis. Of the 93 recovered solutions, 91 executed successfully in our instrumented environment and form the basis of the coverage results. For EDIT-Bench, we compute both whole-file coverage and \emph{diff-region coverage}, where diff-region coverage measures the fraction of edited lines exercised by the tests. We then examine the 39 problems with diff-region coverage below 75\% using structured LLM-assisted labeling (prompt in Appendix~\ref{app:prompt-adequacy}, Figure~\ref{fig:prompt-test-adequacy}) with Claude Sonnet 4.6, and manually verify the cases labeled as genuinely inadequate (see Table~\ref{tab:inadequate-tests} in Appendix~\ref{app:inadequate-manual}). Test counts are reported over all 108 EDIT-Bench problems, while coverage results are reported over the 91 executable problems for which we recovered a passing reference solution.
 
\textbf{Results.}
 
\begin{figure}[t]
  \centering
  \includegraphics[width=0.7\columnwidth]{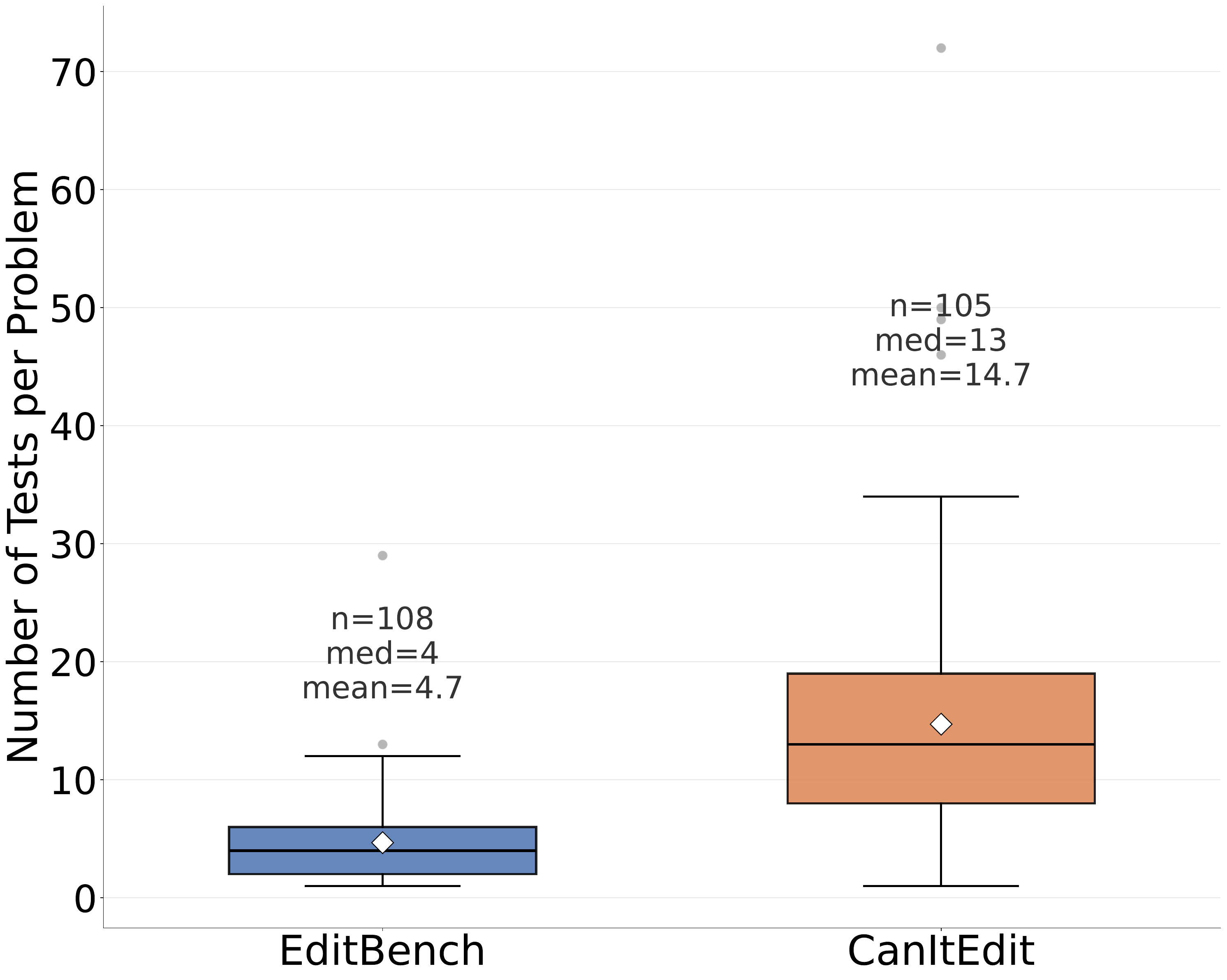}
  \caption{Number of tests per problem.}
  \label{fig:test-counts}
\end{figure}
 
\begin{figure}[t]
  \centering
  \includegraphics[width=0.7\columnwidth]{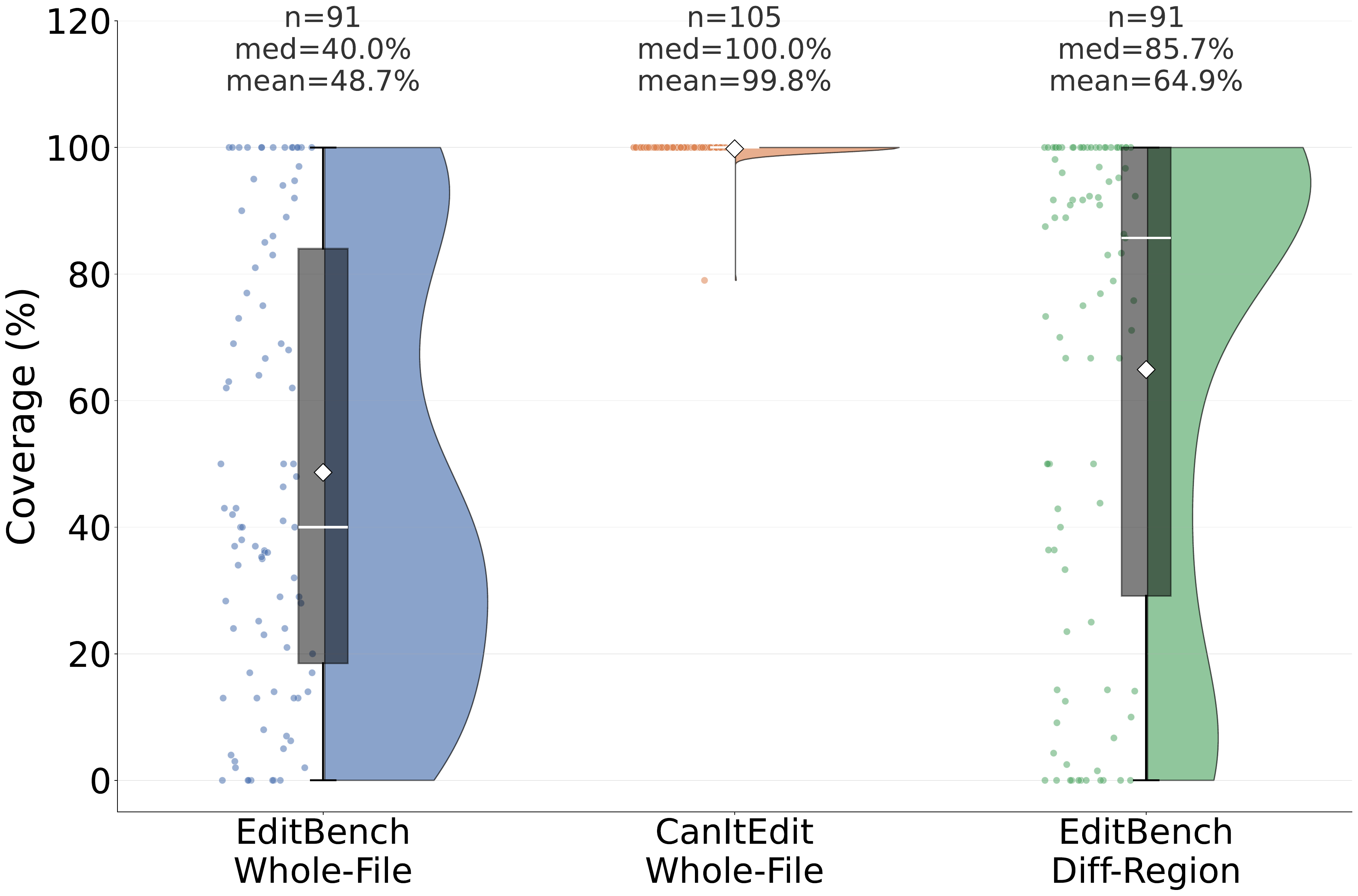}
  \caption{Test coverage distributions of EDIT-Bench and CanItEdit benchmarks.}
  \label{fig:coverage}
\end{figure}
 
\smallskip\noindent\textbf{EDIT-Bench has a median of 4 tests per problem (mean 4.7), compared to CanItEdit's median of 13 (mean 14.7), and the coverage gap is larger still.} Fourteen EDIT-Bench problems rely on a single test as the entire correctness signal (Figure~\ref{fig:test-counts}). CanItEdit achieves near-complete whole-file coverage (median 100.0\%, mean 99.8\%), so we do not report a separate diff-region view for it. EDIT-Bench whole-file coverage is median 40.0\% (mean 48.7\%); for the typical problem, the test suite does not execute even half of the generated solution (Figure~\ref{fig:coverage}). Diff-region coverage improves the picture but does not remove the concern: median rises to 85.7\%, yet the mean remains 64.9\%, 11 problems have 0\% diff-region coverage, and 39 of the 91 executable problems (42.9\%) fall below 75\%. EDIT-Bench sits squarely in the sparse-test regime that prior work has shown can inflate apparent performance under execution-based evaluation~\cite{liu2023evalplus}.

\smallskip\noindent\textbf{Fifty-nine percent of low-coverage EDIT-Bench suites would not detect extraneous modifications outside the requested edit region, and 56\% of all tests scope exclusively to the edited code.} This is particularly concerning given that LLM-powered code editing tools routinely introduce modifications beyond the requested scope in practice~\cite{mozannar2025transformcode}. Of the 39 problems below 75\% diff-region coverage, LLM-assisted reassessment found that 35 (90\%) have low coverage for deliberate reasons: AST or source-text inspection (16), outcome-based testing (14), or heavy mocking (5). Only 4 (10\%) are genuinely inadequate. We verified all four manually and found that their tests detect broken original code but do not exercise the corrected logic; for example, \textit{Problem~87}'s tests check for a \texttt{fill} parameter that the original implementation already satisfies with a constant value, leaving the newly added dynamic colorization and boundary-check logic entirely untested. Even the deliberate strategies that account for the remaining 90\% reveal a deeper problem: AST inspection confirms what the edit adds, similarly, outcome-based testing confirms what the edit produces and mock-heavy tests confirm expected interactions; but none of them confirms what the edit preserves. This makes EDIT-Bench an unreliable proxy for real-world performance as regressions are a routine consequence of software change~\cite{yin2011fixes}. Problem~1 illustrates the pattern: tests count Dropout module instances via \texttt{model.modules()}, so a model that correctly adds Dropout but silently alters convolutional kernel sizes still passes all tests.


\section{Discussion}~\label{sec:discussion}

\smallskip\noindent\textbf{What explains EDIT-Bench's universally unsolved problems?}
Our coverage analysis identified 93 EDIT-Bench problems for which at least one model produced a passing solution. The remaining 15 have no passing solution across any of the 40 models evaluated in the benchmark's published results~\cite{chi2025editbench}. To understand whether these failures reflect model capability limitations or benchmark-level issues, we manually examined outputs from the 12 top-performing models for each problem alongside the original instruction and test suite, and classified each failure by tracing the gap between what the instruction asked, what models produced, and why tests rejected the output (see Table~\ref{tab:unsolved-problems} in Appendix~\ref{app:unsolved}).
 
\smallskip\noindent\textbf{Majority of failures trace to benchmark artifacts rather than model capability boundaries. }Only a minority (4 out of the 15 problems)
reflect genuine model limitations where tests are fair and instructions are reasonable. The clearest model limitation is incomplete semantic understanding of the
requested change. In \textit{Problem~24}, 10 of 12 models add \texttt{async/await} syntax to method signatures and use \texttt{aiofiles} for file operations, but
none replace the synchronous \texttt{requests.get()} call with an async HTTP library. Models understand async syntax but not async semantics. \textit{Problem~9}
reveals a different failure mode: 5 of 12 models unnecessarily remove an existing \texttt{print\_table} method when asked to add a new \texttt{save\_table} method, demonstrating a systematic bias toward modifying existing APIs rather than preserving them. \textit{Problem~94} shows models can get remarkably close without succeeding: the best model fixed 83 of 84 PEP8 violations but overcorrected the final one, removing all trailing newlines instead of normalizing to exactly one.
 
\smallskip\noindent\textbf{The remaining problems fail due to benchmark-level issues spanning test infrastructure bugs, incorrect or overly prescriptive assertions, and unstated environmental requirements.} Several problems are unevaluable regardless of solution quality. In \textit{Problem~39}, the test harness applies dependency mocks after module import has already failed, so a \texttt{ModuleNotFoundError} fires before any assertion executes and all 12 models score 0.0 across all five language variants. \textit{Problem~83} fails because the sandbox never copies a required \texttt{instruction.txt} file to the correct folder, despite models appearing to solve the problem correctly.
 
\smallskip\noindent\textbf{Problems contain test assertions that penalize correct behavior.} \textit{Problem~30}'s \texttt{test\_basic\_functionality} contains mathematical errors, asserting that $5 = 1^2 + 1^2 + 1^2 + 2^2$ when the actual sum is 7. Seven models that correctly return \texttt{False} for these impossible cases are more correct than the expected output. \textit{Problem~53} rejects functionally valid logging implementations through brittle string-matching on source code: one model implemented \texttt{logger.error("Error during run: \%s", e, exc\_info=True)}, which is Python's recommended best practice for exception logging, but the test rejects it because the source code does not contain the exact string \texttt{traceback.format\_exc()}.
 
 
These results reinforce the findings from RQ2. Where RQ2 showed that many test suites verify edit insertion more reliably than edit preservation,
the unsolvable-problem analysis shows that a subset of tests actively penalize correct behavior. EDIT-Bench's apparent difficulty ceiling is substantially inflated by evaluation artifacts rather than fundamental capability boundaries. 
 
\smallskip\noindent\textbf{How much source code do benchmark problems share?}

During the manual analysis of unsolvable problems described above, we noticed that several EDIT-Bench problems present the same source code with different edit instructions. To assess how prevalent this pattern is across both benchmarks, we first embedded each problem's original source code using UniXcoder~\cite{guo2022unixcoder}. We then clustered problems using union-find over pairwise cosine similarities exceeding 0.9, which captures transitive overlap and verified every cluster. We then manually inspected all clustered problems and confirmed that each cluster contains the same or near-identical code assigned different edit instructions.

\smallskip\noindent\textbf{Across both benchmarks, 37 of 213 problems (17\%) share a codebase with at least one other problem.} The issue is far more pronounced in EDIT-Bench: 31 of 108 problems (29\%) cluster into 9 groups, leaving only 86 genuinely distinct code contexts (77 singletons plus 9 unique cluster codebases). The largest cluster contains 11 problems derived from the same Flickr8K dataset loader, differing only in import statements. Other clusters include character-for-character identical files receiving different edit instructions: a stock trading engine appearing three times, a model router class appearing three times, and same-project files from a Telegram bot appearing four times. Several clusters directly compound the domain concentration identified in RQ1: the Flickr8K, LangChain+Ollama, and DuckDuckGo chat API clusters all fall within the ML/AI and chatbot categories that already dominate the benchmark. CanItEdit has much less overlap: 6 of 105 problems (6\%) cluster into 3 pairs, each consisting of byte-level identical code receiving different edit tasks. The lower reuse rate reflects CanItEdit's design as hand-crafted problems rather than repository-sourced samples. 

\smallskip\noindent\textbf{Problem counts in CanItEdit and EDIT-Bench overstate evaluation breadth.} A model evaluated on \textit{Problem~44} (stock trading engine, add commission tracking) and \textit{Problem~97} (same engine, add logging) is tested twice on the same codebase, not on two independent programs. Combined with the domain and language concentration from RQ1, benchmark coverage of real-world editing scenarios is narrower than raw problem counts suggest. For benchmark consumers reporting aggregate pass rates, deduplicating or stratifying by code context would provide a more accurate picture of model capability across genuinely distinct problems.
 
\section{Desiderata for Instructed Code-Editing Benchmarks}~\label{sec:implications}
Instructed code editing underpins some of the most widely deployed AI development tools, yet our survey of over 100 code-related benchmarks yielded only two that target this primitive: CanItEdit and EDIT-Bench. Both concentrate over 90\% of evaluation on a single programming language and leave 31.4\% of real-world edit intent categories entirely unrepresented (RQ1). EDIT-Bench's test suites have measurable oracle-strength gaps affecting 42.9\% of executable problems (RQ2), the majority of its universally unsolved problems trace to benchmark artifacts rather than model limitations, and 29\% of its problems share a codebase with at least one other problem.

While general-purpose guidelines for code benchmark construction exist. For instance, How2Bench~\cite{cao2025how2bench} provides a 55-criterion checklist emphasizing process qualities such as reproducibility, open access, and coverage reporting. These criteria are task-agnostic and focus on \textit{how} to build a benchmark well. Our desiderata address a different question: \textit{what} should an instructed code-editing benchmark specifically contain for its scores to serve as a reliable real-world proxy? Each desideratum below is grounded in a specific finding from our audit.
 
\smallskip\noindent\textbf{\textit{D1: Cover programming languages proportional to ecosystem usage.}}
Both benchmarks are 90\%+ Python. TypeScript, GitHub's most-used language since August 2025~\cite{github2025octoverse}, has zero representation, as do Java, C\#, Go, and Rust. Six languages cover 80\% of new repositories. Editing TypeScript React components or Java Spring services poses different challenges than editing Python scripts, including static type systems and framework conventions that monolingual benchmarks cannot test.
 
\smallskip\noindent\textbf{\textit{D2: Anchor domain distribution of the problems to real editing workloads.}}
EDIT-Bench allocates 36.1\% to AI/ML, a fivefold overrepresentation relative to 7\% prevalence in Copilot Arena~\cite{chi2025copilotarena}. CanItEdit concentrates 68.6\% in algorithmic problem solving, nearly four times its 18\% real-world share. Backend and frontend development, which together account for 46\% of real-world editing activity, are absent from CanItEdit and reach only 24.1\% combined in EDIT-Bench.
 
\smallskip\noindent\textbf{\textit{D3: Include maintenance, documentation, and testing edits.}}
Documentation, test writing, build configuration, and maintenance edits account for 31.4\% of real pull requests~\cite{li2025aidev} but have zero representation in either benchmark. These edit types are structurally different from feature additions and require distinct reasoning about behavior preservation, coverage, and dependency graphs.
 
\smallskip\noindent\textbf{\textit{D4: Verify preservation of existing behaviour (where applicable), not just the edit.}}
56\% of EDIT-Bench suites scope exclusively to edited code, and 59\% of low-coverage suites would not detect modifications outside the edit region. In a domain where 14.8 to 24.4\% of human fixes introduce regressions~\cite{yin2011fixes} and LLMs routinely make extraneous modifications in practice~\cite{mozannar2025transformcode}, checking only ``was the edit made?'' is insufficient. Future benchmarks should include regression-aware suites, report diff-region coverage, and enforce fail-before/pass-after validation. CanItEdit's pipeline, which enforces 100\% coverage and fail-before/pass-after checks, demonstrates this is achievable at benchmark scale.
 
\smallskip\noindent\textbf{\textit{D5: Enforce problem independence and publish quality metadata.}}
29\% of EDIT-Bench problems share a codebase with at least one other problem, and the majority of its universally unsolved problems are benchmark artifacts rather than capability limitations. Future benchmarks should enforce codebase deduplication, validate solvability against reference implementations, and publish problem-level metadata (coverage, solvability, provenance).
 
\smallskip\noindent\textbf{\textit{D6: Representation of emerging development paradigms.}}
Over 1.1 million repositories now use an LLM SDK (+178\% YoY)~\cite{github2025octoverse}, and a majority of surveyed organizations have agents in production~\cite{langchain2025}. Neither benchmark contains problems involving agent orchestration or prompt template management. Future benchmarks should include a forward-looking slice refreshed periodically to track emerging paradigms.

\section{Threats to Validity}~\label{sec:threats}
\paragraph{Internal validity.}
Our domain, edit intent, and ML subcategory classifications rely on LLM-assisted labeling (Claude Sonnet 4.6). While recent work supports LLM reliability for annotating software engineering artifacts~\cite{ahmed2025can,li2025aidev}, misclassifications may affect individual counts; we mitigate this by reporting category-level distributions and manually verifying all cases that directly support a finding. For EDIT-Bench coverage, we used model-generated outputs as reference solutions because the benchmark does not release ground truth. Our coverage measurements therefore reflect the oracle's view of correctness, which is consistent with our goal of assessing the oracle itself.

\paragraph{External validity.}
Our audit covers only two benchmarks. Although these are the only peer-reviewed benchmarks targeting instructed editing that survived our scoping criteria, others may address gaps we identify. CanItEdit is widely adopted and EDIT-Bench was recently published at ICML, so we argue the analysis remains useful; moreover, our methodology and prompts provide a reusable lens for auditing future benchmarks. Further, the reference distributions in RQ1 are proxies, not direct measurements of IDE editing activity: Copilot Arena captures completions, AIDev captures agent-authored PRs, and Octoverse reports repository-level statistics. We use these as the best available empirical anchors, not exact targets. 

\paragraph{Construct validity.}
Statement coverage is a coarse proxy for test adequacy and we agree that branch coverage and mutation scores would provide stronger signals~\cite{liu2025testadequacy} but are not tractable at scale across heterogeneous test frameworks with missing reference solutions. Our preservation assessment (the ``was only the edit made?'' question) relies on LLM-assisted scope labeling, approximating a property ideally measured through mutation testing on non-edited regions. Also, the Conventional Commit taxonomy used for edit intent was designed for commit messages, not instructed editing tasks; some edits may not map cleanly to a single category. We acknowledge these as threats in our study.

\section{Related Work}~\label{sec:related-work}

\smallskip\noindent\textbf{Benchmark quality in code LLM evaluation.} Liu et al.~\cite{liu2023evalplus} showed that augmenting HumanEval's tests lowered pass rates by up to 29\% and reversed model rankings, establishing that sparse tests materially inflate apparent performance. Liu~\cite{liu2025testadequacy} further confirmed that widely used code benchmarks exhibit low branch coverage (74.4\%) and low mutation scores (87.7\%) despite high statement coverage. Cao et al.~\cite{cao2025how2bench} surveyed 572 code benchmarks and found that nearly 70\% lack data quality assurance and only 8.7\% report coverage metadata, proposing a 55-criterion process checklist. Oliva et al.~\cite{oliva2025spice} demonstrated that the kind of problem-level metadata our desideratum D5 calls for is practically achievable at scale, introducing an automated pipeline that labels SWE-bench instances with issue clarity, test coverage, and effort metadata. These studies address code benchmarks broadly and focus on process quality. Our work narrows the focus to instructed code editing and audits content validity: whether the benchmarks test representative edits (RQ1) with oracles strong enough to verify the outcome (RQ2).

\smallskip\noindent\textbf{LLM-assisted code editing.}
Despite its centrality to tools such as GitHub Copilot and Cursor, instructed code editing has received little dedicated evaluation attention; CanItEdit~\cite{cassano2024canitedit} and EDIT-Bench~\cite{chi2025editbench} are the only two peer-reviewed benchmarks targeting this primitive (Section~3). Mozannar et al.~\cite{mozannar2025transformcode} studied LLM-powered editing at scale within Google's internal IDE and identified scope as a recurring failure mode: models frequently modify code beyond the requested region. This directly motivates our RQ2 analysis; our results show that 59\% of low-coverage EDIT-Bench suites would not detect such extraneous changes. Chi et al.~\cite{chi2025copilotarena} demonstrated through Copilot Arena that model rankings under realistic IDE-integrated evaluation diverge from static benchmark rankings. Our RQ1 findings explain why: backend and frontend development account for 46\% of Copilot Arena activity but are largely absent from both benchmarks.

\section{Conclusion}
Instructed code-editing benchmarks serve as a critical proxy to evaluate whether an LLM is ready to ship inside developer-facing AIware such as an IDE. For that proxy to hold, benchmarks must reflect the edits developers actually make and test them with oracles strong enough to back up what a passing score implies. Our audit of CanItEdit and EDIT-Bench shows that neither condition is met. Over 90\% of both benchmarks combined targets Python while TypeScript, GitHub's most-used language, is absent. Problems pertaining to backend and frontend development, which together account for 46\% of real-world editing activity, are largely missing. Documentation, testing, and maintenance edits, which constitute 31.4\% of human PRs, have zero representation. On the oracle side, 42.9\% of executable EDIT-Bench problems fall below 75\% diff-region coverage, 59\% of low-coverage suites would not detect modifications outside the edit region, and the majority of universally unsolved problems trace to benchmark artifacts rather than model limitations. We provide six empirically grounded desiderata addressing language coverage, domain distribution, edit intent breadth, oracle preservation, problem independence, and emerging development paradigms, and release all analysis data and audit artifacts to support the construction of instructed code-editing benchmarks that can serve as reliable real-world proxies.

\section*{Disclaimer}
We used Claude Opus 4.6 and GPT-5.4 for coding assistance and copy-editing. All outputs were manually verified by the authors. This usage complies with both IEEE and ACM policies on generative AI in publications.


\bibliographystyle{ACM-Reference-Format}
\bibliography{main}

\newpage

\appendix

\section{Classification Prompts}
\label{app:prompts}

\begin{figure}[t]
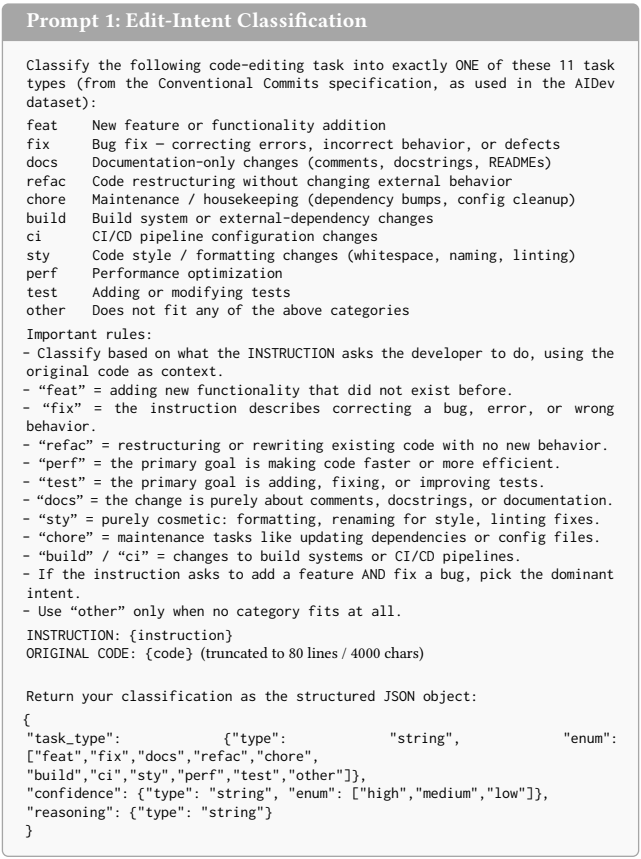

\centering
\begin{tcolorbox}[
  colback=gray!5,
  colframe=gray!70,
  fonttitle=\bfseries\small,
  title=Prompt 1: Edit-Intent Classification,
  boxrule=0.5pt,
  arc=2pt,
  left=6pt, right=6pt, top=4pt, bottom=4pt,
  fontupper=\scriptsize\ttfamily
]
Classify the following code-editing task into exactly 
ONE of these 11 task types (from the Conventional 
Commits specification, as used in the AIDev dataset):

\smallskip
\begin{tabular}{@{}ll@{}}
feat  & New feature or functionality addition \\
fix   & Bug fix --- correcting errors, incorrect 
        behavior, or defects \\
docs  & Documentation-only changes (comments, 
        docstrings, READMEs) \\
refac & Code restructuring without changing external 
        behavior \\
chore & Maintenance / housekeeping (dependency bumps, 
        config cleanup) \\
build & Build system or external-dependency changes \\
ci    & CI/CD pipeline configuration changes \\
sty   & Code style / formatting changes (whitespace, 
        naming, linting) \\
perf  & Performance optimization \\
test  & Adding or modifying tests \\
other & Does not fit any of the above categories \\
\end{tabular}

\smallskip
Important rules:\\
- Classify based on what the INSTRUCTION asks the 
  developer to do, using the original code as context.\\
- ``feat'' = adding new functionality that did not 
  exist before.\\
- ``fix'' = the instruction describes correcting a 
  bug, error, or wrong behavior.\\
- ``refac'' = restructuring or rewriting existing code 
  with no new behavior.\\
- ``perf'' = the primary goal is making code faster or 
  more efficient.\\
- ``test'' = the primary goal is adding, fixing, or 
  improving tests.\\
- ``docs'' = the change is purely about comments, 
  docstrings, or documentation.\\
- ``sty'' = purely cosmetic: formatting, renaming for 
  style, linting fixes.\\
- ``chore'' = maintenance tasks like updating 
  dependencies or config files.\\
- ``build'' / ``ci'' = changes to build systems or 
  CI/CD pipelines.\\
- If the instruction asks to add a feature AND fix a 
  bug, pick the dominant intent.\\
- Use ``other'' only when no category fits at all.

\smallskip

INSTRUCTION: \{instruction\}\\
ORIGINAL CODE: \{code\} 
\textrm{\scriptsize(truncated to 80 lines / 4000 chars)}\\

\smallskip
Return your classification as the structured JSON object:

\smallskip
\{\\
\quad"task\_type": \{"type": "string", "enum": 
  ["feat","fix","docs","refac","chore",\\
\quad\quad"build","ci","sty","perf","test","other"]\},\\
\quad"confidence": \{"type": "string", "enum": 
  ["high","medium","low"]\},\\
\quad"reasoning": \{"type": "string"\}\\
\}
\end{tcolorbox}
\caption{Prompt template for classifying benchmark tasks 
into Conventional Commit categories (RQ1).}
\label{fig:prompt-task-type}
\end{figure}

\begin{figure}[t]
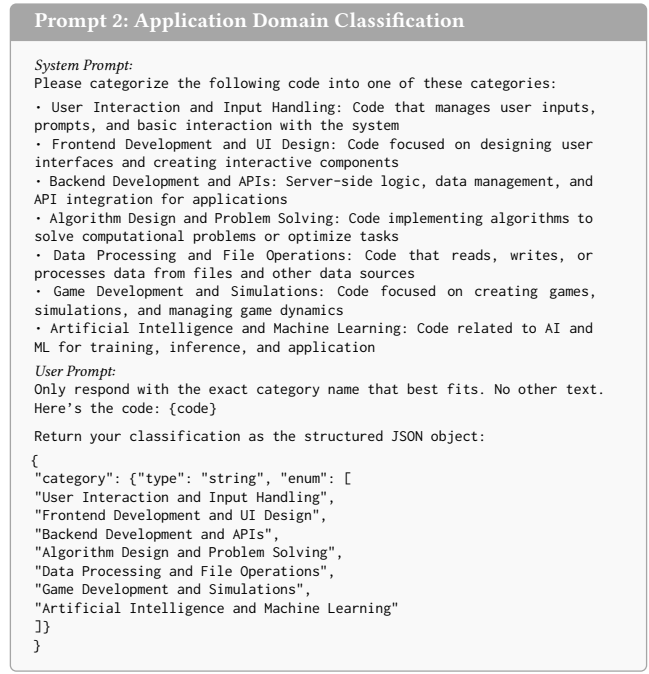

\centering
\begin{tcolorbox}[
  colback=gray!5,
  colframe=gray!70,
  fonttitle=\bfseries\small,
  title=Prompt 2: Application Domain Classification,
  boxrule=0.5pt,
  arc=2pt,
  left=6pt, right=6pt, top=4pt, bottom=4pt,
  fontupper=\scriptsize\ttfamily
]
\textrm{\scriptsize\textit{System Prompt:}}\\
Please categorize the following code into one of these 
categories:

\smallskip
\begin{tabular}{@{}p{0.95\linewidth}@{}}
\textbullet\ User Interaction and Input Handling: 
  Code that manages user inputs, prompts, and basic 
  interaction with the system \\
\textbullet\ Frontend Development and UI Design: 
  Code focused on designing user interfaces and 
  creating interactive components \\
\textbullet\ Backend Development and APIs: 
  Server-side logic, data management, and API 
  integration for applications \\
\textbullet\ Algorithm Design and Problem Solving: 
  Code implementing algorithms to solve computational 
  problems or optimize tasks \\
\textbullet\ Data Processing and File Operations: 
  Code that reads, writes, or processes data from 
  files and other data sources \\
\textbullet\ Game Development and Simulations: 
  Code focused on creating games, simulations, and 
  managing game dynamics \\
\textbullet\ Artificial Intelligence and Machine 
  Learning: Code related to AI and ML for training, 
  inference, and application \\
\end{tabular}

\smallskip
\textrm{\scriptsize\textit{User Prompt:}}\\
Only respond with the exact category name that best 
fits. No other text.\\
Here's the code: \{code\}

\smallskip

\smallskip
Return your classification as the structured JSON object:

\smallskip
\{\\
\quad"category": \{"type": "string", "enum": [\\
\quad\quad"User Interaction and Input Handling",\\
\quad\quad"Frontend Development and UI Design",\\
\quad\quad"Backend Development and APIs",\\
\quad\quad"Algorithm Design and Problem Solving",\\
\quad\quad"Data Processing and File Operations",\\
\quad\quad"Game Development and Simulations",\\
\quad\quad"Artificial Intelligence and Machine 
  Learning"\\
\quad]\}\\
\}
\end{tcolorbox}
\caption{Prompt template for classifying benchmark tasks 
into application domains (RQ1). Categories are aligned 
with those reported by Copilot 
Arena~\cite{chi2025copilotarena}.}
\label{fig:prompt-app-domain}
\end{figure}

This appendix documents the LLM prompts used for the three classification tasks in our audit. All classifications were performed using Claude Sonnet 4.6. Section~\ref{app:prompt-intent} presents the edit-intent prompt used in RQ1 (Figure~\ref{fig:prompt-task-type}). Section~\ref{app:prompt-domain} presents the application-domain prompt used in RQ1 (Figure~\ref{fig:prompt-app-domain}). Section~\ref{app:prompt-ml} presents the AI/ML subcategory prompt used to produce Table~\ref{tab:ml-subcategory} in RQ1 (Figure~\ref{fig:prompt-ml-subtype}).

\subsection{Edit-Intent Classification}
\label{app:prompt-intent}

The following prompt classifies each benchmark problem into one of eleven Conventional Commit categories, matching the taxonomy used by the AIDev dataset~\cite{li2025aidev}. Results are reported in Figure~\ref{fig:edit-intent}.

\subsection{Application-Domain Classification}
\label{app:prompt-domain}

The following prompt classifies each benchmark problem into one of seven application domains aligned with those reported by Copilot Arena~\cite{chi2025copilotarena}. Results are reported in Figure~\ref{fig:app-domain}.

\subsection{AI/ML Subcategory Classification}
\label{app:prompt-ml}

The following prompt further classifies problems tagged as AI/ML into three subcategories to assess representation of emerging development paradigms. Results are reported in Table~\ref{tab:ml-subcategory}.

\begin{figure}[t]
\centering
\begin{tcolorbox}[
  colback=gray!5,
  colframe=gray!70,
  fonttitle=\bfseries\small,
  title=Prompt 3: AI/ML Subcategory Classification),
  boxrule=0.5pt,
  arc=2pt,
  left=6pt, right=6pt, top=4pt, bottom=4pt,
  fontupper=\scriptsize\ttfamily
]
You are classifying code-editing problems by the TYPE 
of AI/ML work involved.

Given a code-editing task (instruction + code context), 
classify it into EXACTLY ONE of these categories:

\smallskip
\textbf{AI-NATIVE-APPLICATION-DEV}: Code that USES 
AI/ML models as services or components within a larger 
application. This includes:\\
- Calling LLM APIs (OpenAI, Anthropic, Ollama, 
  HuggingFace Inference API, etc.)\\
- Using LangChain, LlamaIndex, or other LLM 
  orchestration frameworks\\
- Building RAG pipelines, prompt templates, or chat 
  interfaces that consume AI APIs\\
- Configuring or deploying pre-trained models as 
  services (not training them)\\
- Agent framework configuration and tool-use setup\\
- Building applications ON TOP of AI capabilities

\smallskip
\textbf{TRADITIONAL-ML-DEV}: Code that BUILDS, TRAINS, 
or EVALUATES ML/AI models. This includes:\\
- Defining model architectures (CNN, RNN, Transformer 
  layers)\\
- Training loops, loss functions, optimizers, learning 
  rate schedules\\
- Data preprocessing FOR model training (feature 
  engineering, train/val splits)\\
- Model evaluation metrics and validation\\
- Hyperparameter tuning\\
- Fine-tuning or adapter training (LoRA, PEFT) --- the 
  training side, not inference/serving

\smallskip
\textbf{ML-ADJACENT-TOOLING}: Code that works WITH ML 
outputs but is not itself model-building or 
API-calling. This includes:\\
- Visualization of model results (heatmaps, confusion 
  matrices, plots)\\
- Data analysis and exploration using ML library 
  outputs\\
- Statistical analysis using scipy/sklearn for non-ML 
  purposes\\
- File/data management utilities involving model 
  artifacts

\smallskip

INSTRUCTION: \{instruction\}\\
CODE CONTEXT: \{code\} 
\textrm{\scriptsize(first 800 chars)}\\

\smallskip
Return your classification as the structured JSON object:

\smallskip
\{\\
\quad"classification": \{"type": "string", "enum": 
  ["AI-NATIVE-APPLICATION-DEV",\\
\quad\quad"TRADITIONAL-ML-DEV",
  "ML-ADJACENT-TOOLING"]\},\\
\quad"confidence": \{"type": "string", "enum": 
  ["high","medium","low"]\},\\
\quad"evidence": \{"type": "string"\}\\
\}
\end{tcolorbox}
\caption{Prompt template for classifying AI/ML problems 
into subcategories (RQ1).}
\label{fig:prompt-ml-subtype}
\end{figure}

\section{Test Adequacy Assessment}
\label{app:test-adequacy}

This appendix documents the prompt and manual verification results for the test adequacy analysis in RQ2. Section~\ref{app:prompt-adequacy} presents the LLM prompt used to assess whether low-coverage test suites are genuinely inadequate or explained by the testing strategy. Section~\ref{app:inadequate-manual} reports the manual analysis of the four problems identified as genuinely inadequate.

\clearpage
\subsection{Test Adequacy Assessment Prompt}
\label{app:prompt-adequacy}

The following prompt was applied to the 39 EDIT-Bench problems with diff-region coverage below 75\%. Of these, 35 were classified as having low coverage for deliberate reasons (AST/source inspection, outcome-based testing, or heavy mocking) and 4 as genuinely inadequate. All four genuinely inadequate cases were manually verified (Table~\ref{tab:inadequate-tests}).

\begin{figure}[t]
\centering
\begin{tcolorbox}[
  colback=gray!5,
  colframe=gray!70,
  fonttitle=\bfseries\small,
  title=Prompt 4: Test Suite Adequacy Assessment,
  boxrule=0.5pt,
  arc=2pt,
  left=6pt, right=6pt, top=4pt, bottom=4pt,
  fontupper=\scriptsize\ttfamily
]
You are an expert software testing analyst. Assess 
whether the test suite for a code editing problem 
adequately captures the INTENT of the edit instruction.

\smallskip
\textbf{Context}\\
A developer was given code and asked to make a specific 
edit. A test suite was written to verify the edit. You 
must assess whether the tests actually verify the edit 
intent, or only check surface-level properties.

\smallskip
\textbf{Measured Coverage Data}\\
- Whole-file line coverage: \{whole\_file\_coverage\}\%\\
- Diff-region coverage (only the changed lines): 
  \{diff\_coverage\}\%\\
- Number of diff lines (additions + removals): 
  \{diff\_lines\}\\
The diff coverage is BELOW 75\%. Your job is to 
determine whether this is a real problem or explained 
by the testing approach.

\smallskip

Edit Instruction: \{instruction\}\\
Original Code (before the edit): \{original\_code\}\\
Highlighted Section (the part to be changed): 
\{highlighted\_code\}\\
Test Suite: \{test\_code\}\\

\smallskip
\textbf{Assessment Dimensions}

\smallskip
\begin{tabular}{@{}rp{0.88\linewidth}@{}}
1. & \textbf{Edit-specificity}: Do tests specifically 
     verify the changes described in the instruction? \\
2. & \textbf{Discrimination power}: Would tests FAIL if 
     original\_code was submitted unchanged? Could tests 
     be FOOLED by a superficially similar but incorrect 
     edit? \\
3. & \textbf{Testing approach}: BEHAVIORAL (exercises 
     runtime), STRUCTURAL (inspects code text/AST), or 
     MIXED. \\
4. & \textbf{Low coverage reason}: Why is diff coverage 
     at \{diff\_coverage\}\%? One of: 
     \textit{ast\_source\_analysis}, 
     \textit{heavy\_mocking}, 
     \textit{outcome\_based\_testing}, or 
     \textit{genuinely\_inadequate}. \\
5. & \textbf{Detection of unwanted changes}: If a model 
     correctly makes the requested edit BUT ALSO modifies 
     other parts of the code that should remain unchanged, 
     would the tests catch it? One of: \textit{yes}, 
     \textit{partially}, \textit{no}. \\
6. & \textbf{Test scope}: Does the test suite only verify 
     the edit itself (\textit{edit\_only}), also check 
     surrounding behavior 
     (\textit{edit\_plus\_context}), or exercise the 
     whole file broadly (\textit{whole\_file})? \\
\end{tabular}

\smallskip
Provide your assessment as structured JSON:

\smallskip
\{\\
\quad"adequacy\_rating": \{"type": "string"\},\\
\quad"edit\_specificity": \{"type": "string"\},\\
\quad"discrimination\_power": \{\\
\quad\quad"catches\_missing\_edit": \{"type": "boolean"\},\\
\quad\quad"fooled\_by\_wrong\_edit": \{"type": 
  "boolean"\}\\
\quad\},\\
\quad"testing\_approach": \{"type": "string", "enum": 
  ["behavioral","structural","mixed"]\},\\
\quad"low\_coverage\_reason": \{"type": "string", "enum": 
  ["ast\_source\_analysis",\\
\quad\quad"heavy\_mocking","outcome\_based\_testing",
  "genuinely\_inadequate"]\},\\
\quad"is\_low\_coverage\_a\_real\_problem": \{"type": 
  "boolean"\},\\
\quad"detects\_unwanted\_changes": \{"type": "string", 
  "enum": ["yes","partially","no"]\},\\
\quad"detects\_unwanted\_changes\_explanation": 
  \{"type": "string"\},\\
\quad"test\_scope": \{"type": "string", "enum": 
  ["edit\_only","edit\_plus\_context",
  "whole\_file"]\},\\
\quad"rationale": \{"type": "string"\}\\
\}
\end{tcolorbox}
\caption{Prompt template for assessing test suite 
adequacy against edit intent (RQ2).}
\label{fig:prompt-test-adequacy}
\end{figure}

\subsection{Manual Verification of Inadequate Test Suites}
\label{app:inadequate-manual}

Table~\ref{tab:inadequate-tests} presents the manual analysis of the four EDIT-Bench problems that the LLM-assisted assessment in Section~\ref{app:prompt-adequacy} flagged as genuinely inadequate. For each problem, we traced what the edit changes, why the tests fail to exercise that change, and the resulting consequence for evaluation validity.

\begin{table*}[t]
\centering
\caption{Manual analysis of four EDIT-Bench problems identified as having genuinely inadequate test suites (diff-region coverage below 75\%).}
\label{tab:inadequate-tests}
\small
\begin{tabular}{@{}cp{4.8cm}p{4.8cm}p{4.2cm}@{}}
\toprule
\textbf{Problem} & \textbf{What the edit changes} & \textbf{Why the tests are inadequate} & \textbf{Consequence} \\
\midrule
104 & Fixes \texttt{add\_edge} logic, re-indents \texttt{remove\_vertex} and \texttt{remove\_edge}, and corrects print-string formatting. & Tests detect failures only because the original file contains a syntax/import error, not because they verify the repaired logic. None of the fixed methods are exercised. & Suite confirms the original is malformed but provides no confidence the fix is functionally correct. \\
\addlinespace
29 & Modifies pandas column-processing logic in \texttt{visualize\_results\_generic}. & The test intended to call the function with sample DataFrames is overwritten by a second test with the same name, leaving only a shallow existence check. & Modified column-handling code is never validated. \\
\addlinespace
52 & Rewrites DataFrame construction from scalar-based to list-per-column format. & Tests inspect source text and reconstruct a DataFrame in the harness rather than executing the implementation. Any non-list-of-dicts pattern is accepted as corrected. & Original broken version can still appear to pass; changed code is never executed. \\
\addlinespace
87 & Adds dynamic square colorization and a boundary check preventing squares from disappearing before the canvas is drawn. & One test checks for a \texttt{fill} parameter, but the original constant blue already satisfies it. The movement test covers only simple in-bounds moves and never triggers the new boundary logic. & Neither new behavior (dynamic color, boundary check) is exercised by the suite. \\
\bottomrule
\end{tabular}
\end{table*}

\section{Universally Unsolved Problem Analysis}
\label{app:unsolved}

Section~\ref{sec:discussion} reported that 11 of 15 universally unsolved EDIT-Bench problems trace to benchmark artifacts rather than model limitations. Table~\ref{tab:unsolved-problems} presents the complete classification. For each problem, one author examined outputs from the 12 top-performing models alongside the original instruction and test suite, and classified the failure by tracing the gap between what the instruction asked, what models produced, and why tests rejected the output. Problems are grouped by primary failure cause.

\begin{table*}[t]
\centering
\caption{Classification of universally unsolved EDIT-Bench problems. }
\label{tab:unsolved-problems}
\small
\begin{tabular}{@{}clp{10.5cm}@{}}
\toprule
\textbf{ID} & \textbf{Primary Cause} & \textbf{Summary} \\
\midrule
\multicolumn{3}{@{}l}{\textit{Benchmark Artifacts}} \\
\addlinespace
53 & Test Unfairness & Tests use brittle string-matching on source code rather than validating behavior. A model implementing Python's recommended \texttt{logger.error(..., exc\_info=True)} is rejected because the source does not contain the exact string \texttt{traceback.format\_exc()}. \\
\addlinespace
30 & Invalid Test Logic & \texttt{test\_basic\_functionality} contains mathematical errors (e.g., asserts $5 = 1^2+1^2+1^2+2^2$ when the actual sum is 7). Seven models correctly return \texttt{False} for these impossible cases but are penalized. \\
\addlinespace
8 & Test Bug + Underspecification & Instruction is a title fragment with no actionable guidance. Tests require module importability (unstated) and contain a mathematical error rejecting the valid decomposition $4 = 1^2+1^2+1^2+1^2$. \\
\addlinespace
126 & Fragile Mocking & All 12 models correctly implement dictionary-based clustering, but extremely complex mocking (\texttt{patch.dict}, nested \texttt{try/except}, \texttt{patch.object}) breaks with minor implementation variations. Models score 0.167--0.333 despite correct behavior. \\
\addlinespace
39 & Infrastructure Bug & Test harness applies dependency mocks \emph{after} module import has already failed. \texttt{ModuleNotFoundError} fires before any assertion executes; all 12 models score 0.0 across all 5 language variants. \\
\addlinespace
83 & Infrastructure Bug & Sandbox never copies a required \texttt{instruction.txt} file to the correct folder. Models appear to solve the problem correctly but cannot be evaluated. \\
\addlinespace
6 & Unstated Requirement & Models correctly fix indentation errors as instructed, but tests require module importability. The code contains blocking \texttt{input()} calls designed for interactive use; the importability constraint is never mentioned in the instruction. \\
\addlinespace
32 & Instruction Ambiguity & Chinese instruction uses ``\begin{CJK}{UTF8}{gbsn}功能\end{CJK}'' which is ambiguous between ``a function'' (callable) and ``functionality'' (behavior). All 12 models unanimously add functionality to the existing loop; tests expect a new callable. \\
\addlinespace
\midrule
\multicolumn{3}{@{}l}{\textit{Genuine Model Limitations}} \\
\addlinespace
24 & Incomplete Async Conversion & 10 of 12 models add \texttt{async/await} syntax but fail to replace the synchronous \texttt{requests.get()} with an async HTTP library. Models understand async syntax but not async semantics. \\
\addlinespace
9 & Over-aggressive API Modification & 5 of 12 models unnecessarily remove the existing \texttt{print\_table} method when asked to add a new \texttt{save\_table} method, demonstrating a bias toward modifying existing APIs rather than preserving them. \\
\addlinespace
94 & Incomplete Task Execution & Best model fixed 83 of 84 PEP8 violations (98.8\%) but overcorrected the final one, removing all trailing newlines instead of normalizing to exactly one. PEP8 compliance is binary. \\
\addlinespace
50 & Structural Refactoring Failure & Models interpret a vague instruction as requiring only an algorithm change, performing minimal line-level edits instead of fixing a critical definition-usage mismatch (\texttt{is\_int} defined at module level but called as \texttt{IntegerToken.is\_int()}). \\
\bottomrule
\end{tabular}
\end{table*}

\end{document}